\documentclass[aps,prb,reprint,twocolumn,superscriptaddress,longbibliography]{revtex4-2}

\usepackage[utf8]{inputenc}
\usepackage{graphicx}
\usepackage{amsmath,amssymb}
\usepackage{bm}
\usepackage{xcolor}
\usepackage{booktabs}
\usepackage[colorlinks=true,citecolor=blue,linkcolor=blue,urlcolor=blue]{hyperref}

\newcommand{\vk}{\bm{k}}

\newcommand{\qc}{q_c}
\begin{document}

\title{Chirality rules for optical vortices in rhombohedral graphene multilayers}

\author{Felipe~P\'erez Riffo}
\affiliation{Departamento de F\'isica, Universidad T\'ecnica Federico Santa Mar\'ia, Valpara\'iso, Chile}

\author{Mat\'ias~Castro Schnaidt}
\affiliation{Departamento de F\'isica, Universidad T\'ecnica Federico Santa Mar\'ia, Valpara\'iso, Chile}

\author{Juan M. Florez}
\affiliation{Facultad de Ingenier\'ia y Arquitectura, Universidad Central de Chile, Santiago, Chile}

\author{E.~Su\'arez Morell}
\email{eric.suarez@usm.cl}
\affiliation{Departamento de F\'isica, Universidad T\'ecnica Federico Santa Mar\'ia, Valpara\'iso, Chile}

\date{\today}
\begin{abstract}
Optical zeros are momenta where the interband dipole vanishes.  The dipole
phase winds around each of them by an integer, so they are vortices of the
resonant optical response with a conserved charge.  We show that in
rhombohedral graphene multilayers one integer, the stacking chirality $N$,
fixes their constellation near the $K$ valley:\ the total charge per linear
polarization is $N-1$ and is conserved under trigonal warping and interlayer
bias; for $N\ge3$ the $K$ point stays an exact optical zero, the Bernal
bilayer being the only exception; warping places the remaining charge on
satellites whose radii scale linearly with the warping; and the charge pinned
at $K$ is bias independent for $N\ge4$.  A quantized interband character
counts the enclosed charge as an integer staircase with steps at these radii,
and an interlayer bias drives a satellite through the protected $K$ vortex
in the trilayer.  The rules survive the remote hoppings of the full
tight-binding model and, being integer charges, the corrections that reshape
the bands.  The same integer sets the magnitude of the response:\ the
band-edge shift current grows linearly with $N$.
\end{abstract}

\maketitle

\section{Introduction}

The bulk photovoltaic effect (BPVE) converts light into dc current in a single
noncentrosymmetric crystal, without a junction.  Its dominant intrinsic
mechanism, the shift current, is set by the quantum geometry of the Bloch
states~\cite{Ahn2022,JiangHolderYan2025,Verma2026}:\ photoexcited carriers are
displaced in real space by the shift vector, a gauge-invariant combination of
Berry connections of the valence and conduction
bands~\cite{vonBaltz1981,Sipe2000,Young2012,Cook2017}.  The effect has
recently been measured directly in two-dimensional ferroelectrics and
semiconductors~\cite{ChangSnS2023,NakamuraShift2024}.

Within this picture, much attention has turned to \emph{optical zeros}:\
momenta where the interband dipole of a given polarization
vanishes~\cite{Zhu2024,Alexandradinata2024}.  At an optical zero the
transition is dark, the phase of the dipole winds by an integer, and the shift
vector diverges.  Optical zeros are therefore the vortices of the resonant
response, and their windings quantize contributions to the shift
current~\cite{Alexandradinata2024,Avdoshkin2025}.  Vortex textures of the
shift photocurrent have been discussed in polar and moir\'e two-dimensional
materials~\cite{npj2DVortices2025,PNASVortexCrystals2023}.

Chirally (rhombohedrally) stacked graphene multilayers are a natural
platform for this physics.  Their low-energy bands carry a pseudospin winding
equal to the layer number $N$~\cite{Min2008,McCannKoshino2013}, which
dictates optical selection rules~\cite{Cao2018,Zhang2018}; the
trigonal-warping hopping $\gamma_3$ splits the band touching into satellite
Dirac cones~\cite{KoshinoMcCann2009}; and a perpendicular electric field opens
a widely tunable gap~\cite{ZhangBLGgap2009,ZanABC2024}.  The same stacks host
gate-tunable correlated and topological
phases~\cite{LuFQAH2024,HanPenta2024,ChoiRhombo2025,ZhangNlayer2025}, and
sizable shift currents have been predicted for twisted and rhombohedral
stacks~\cite{TBGShift2022,RTG2024,SuperlatticeBLG2025}.

Our interest in optical zeros is concrete.  In biased bilayer and trilayer
graphene, we found that the band-edge shift current is governed to a large
extent by the zeros that sit at or near the band edge, where the shift vector
diverges and the integrand changes sign; furthermore, the current itself can be
decomposed into the integer charges of these
zeros~\cite{SelectionRule2026,CastroSchnaidt2026}.  Knowing how many zeros a
stack has, where they sit, and which couplings move them is therefore a
direct route to its photocurrent.  

The formalism that counts the windings of
optical zeros is established~\cite{Alexandradinata2024,Zhu2024}, but it has
been applied case by case: the zeros of a given material have not been
derived analytically, nor for a family of structures in which the answer could be tested against a tunable integer.  

The windings of the dipole of
chiral fermions have also been exploited for vortex-enhanced photovoltaic
currents in disordered Dirac and Weyl materials, including the continuum
model of rhombohedral graphene~\cite{Sukhachov2026}.  In that work, the authors keep the
isotropic chiral model, in which the whole vortex charge sits at $K$, and
trigonal warping enters only as a perturbation of the current.  Here trigonal
warping plays the opposite role:\ it is the term that splits the charge at $K$
into a constellation of zeros at finite momentum, whose positions the
perpendicular field then tunes.  It also remains to
be shown how the band-edge shift current, which trigonal warping switches on in these stacks~\cite{SelectionRule2026}, grows with $N$.

Both questions are answered by the stacking chirality.  The effective coupling between
the outer layers is $\bar h_N\propto\xi^{N}$, with $\xi=q_x+iq_y$ the momentum
measured from the $K$ valley.  Its phase winds $N$ times around $K$, which
fixes the vortex charge; its modulus steepens with $N$ on the band-edge ring (the ring of minimal direct gap, at the rim of the gapped Mexican-hat bands, on which the band-edge transitions concentrate), which
sets the size of the photocurrent.  This integer organizes the paper.

On the phase side, we chart the constellation of optical zeros near $K$
across the family (monolayer, Bernal (AB) bilayer, rhombohedral (ABC) trilayer, and the
generic $N$-layer) and derive four analytic rules for their number,
positions, and windings (Sec.~\ref{sec:rules}).  The central one is a
protection rule:\ for $N\ge3$ the $K$ point remains an \emph{exact} dark
transition for any warping and bias, and the Bernal bilayer is the only member
whose vortex is expelled.

The other three rules state that the total charge, $N-1$,
inherited from the chiral matrix element, is conserved under warping and
bias, which only redistribute it among zeros of integer charge;
place every satellite (an optical zero that warping or bias detaches from $K$, not to be confused with the satellite Dirac cones of the spectrum) on a single warping scale with a bias-tunable fine
structure; and fix the charge pinned at $K$.  All four rules are checked against
full tight-binding (TB) calculations.

The quantized interband character $\chi=\nu-F$
(optical vorticity minus interband Berry flux), which we build from the circulation identity of Ref.~\cite{Zhu2024} and evaluate on
disks of growing radius, counts this constellation as an integer staircase whose steps sit at
the analytic radii, while the band Chern numbers are unaffected by every
parameter that reshapes it (Sec.~\ref{sec:counting}).  The constellation
leaves its imprint on the shift-current integrand, whose nodal structure is
anchored on the optical zeros (Sec.~\ref{sec:imprint}).  We also show that
the band-edge peak $|\sigma|_{\rm pk}$ of the shift conductivity $\sigma^{yyy}$ grows linearly with layer number, $|\sigma|_{\rm pk}\propto N$, the chirality
entering through the interband dipole (Sec.~\ref{sec:Nlaw} and
Appendix~\ref{app:Nlaw}).

Section~\ref{sec:models} defines the models and observables,
Sec.~\ref{sec:Nlayer} tests the rules against the remote hoppings, where the
competition between bias and the vertical hopping $\gamma_2$ produces a
gate-tunable vortex merger, and Sec.~\ref{sec:discussion} discusses the scope of the results.

\section{Models and methods}
\label{sec:models}

\subsection{Tight-binding models}

We consider rhombohedral $N$-layer graphene with basis
$(A_1,B_1,\dots,A_N,B_N)$ and in-plane structure factor
\begin{equation}
f(\vk)=e^{i k_y a/\sqrt3}+2\,e^{-i k_y a/(2\sqrt3)}\cos(k_x a/2),
\end{equation}
with $a=\sqrt3\,a_{cc}$, $a_{cc}=1.42$~\AA, and $f(K)=0$ at
$K=(4\pi/3a,0)$.  The couplings are the intralayer hopping $\gamma_0$
($A_l$--$B_l$, amplitude $-\gamma_0 f$), the interlayer dimer $\gamma_1$
($B_l$--$A_{l+1}$), the trigonal-warping skew hopping $\gamma_3$
($A_l$--$B_{l+1}$, amplitude $-\gamma_3 f^{*}$). The remote
couplings, namely $\gamma_4$ ($A_l$--$A_{l+1}$ and $B_l$--$B_{l+1}$, amplitude
$\gamma_4 f$), the vertical hopping $\gamma_2$ ($A_l$--$B_{l+2}$, which in the
trilayer links the two low-energy sites $A_1$ and $B_3$ directly), and the energy $\Delta'$ of
the dimer sites relative to the others, are set to zero except where stated
(Sec.~\ref{sec:Nlayer}). Unless stated otherwise $\gamma_0=3.16$, $\gamma_1=0.39$, and
$\gamma_3=0.30$~eV (for the bilayer, $\gamma_1=0.381$ and $\gamma_3=0.38$~eV).

An interlayer bias enters as a linear potential ladder from $+U$
(layer~1) to $-U$ (layer~$N$).  It is essential throughout:\ pristine AB and
ABC stacks are centrosymmetric and carry no shift
current~\cite{FewLayer2026}.  For the Bernal bilayer we keep the
conventional assignment, layer~1 at $-\delta$ and layer~2 at $+\delta$, which
orients the field opposite to the rhombohedral ladder.  The choice is only a
label:\ reversing the bias moves no zero and changes no magnitude, but flips
the sign of every winding and of the shift current together (Rule~1 of Sec.~\ref{sec:rules}).  Signs
quoted for the bilayer therefore carry the opposite sign convention:\ its
vortex charges and staircase steps appear as $-1$ where the rhombohedral
stacks give $+1$.

The two low-energy sites are $A_1$ and $B_N$.  Because the skew hopping
connects $A_l$ to $B_{l+1}$ only, it bridges them directly in the bilayer
($A_1$--$B_2$) but \emph{never for $N\ge3$}.  This structural fact underlies
the protection rule of Sec.~\ref{sec:rules}.

\subsection{Optical zeros, windings, and shift current}

The central object is the interband velocity matrix element
$v^{a}_{nm}(\vk)=\langle\psi_n|\partial H/\partial k_a|\psi_m\rangle$ between
the valence and conduction bands $(v,c)$ that bound the gap, the amplitude
for a vertical transition driven by light polarized along $a$ (the Cartesian indices $a,b,c\in\{x,y\}$).  An
\emph{optical zero} is a momentum at which $v^{a}_{vc}$ vanishes, so that the
transition there is dark for that polarization.  Since $v^{a}_{vc}$ is a
complex field, the winding of its phase on a small loop around the zero,
$w=\tfrac{1}{2\pi}\oint d\,\mathrm{arg}\,v^{a}_{vc}$, is an integer that no
smooth deformation can change:\ the \emph{vortex charge} of the zero, which
the rules of Sec.~\ref{sec:rules} fix and the staircase of
Sec.~\ref{sec:counting} counts.  Numerically we locate the zeros as minima of
the gauge-invariant $|v^{a}_{vc}|$ and read their charges off small loops in a
smooth reference gauge.
The charge is gauge invariant, since a smooth, single-valued change of band
gauge, $\psi_n\to e^{i\phi_n(\vk)}\psi_n$, multiplies $v^{a}_{vc}$ by
$e^{i(\phi_c-\phi_v)}$ and adds no circulation (Appendix~\ref{app:census}).

The stacks are $C_3$ symmetric with a vertical mirror, which leaves a
single independent component of the third-rank conductivity tensor $\sigma^{abc}$, defined by $j^{a}=\sigma^{abc}(\omega)E^{b}E^{c}$ for the dc current density $j^{a}$ generated by an optical field $\bm E$ of frequency $\omega$.  The
mirror forbids every component with an odd number of $x$ indices (in
particular $\sigma^{xxx}=0$), and $C_3$ ties the survivors together,
$\sigma^{yyy}=-\sigma^{yxx}=-\sigma^{xxy}$.  We therefore work throughout with
$\sigma^{yyy}$.

The sheet conductivity is set by the transition dipole
$r^{y}_{cv}=v^{y}_{cv}/(i\omega_{cv})$, where $\omega_{nm}=E_n-E_m$ is a transition energy ($\hbar=1$ in $\omega_{nm}$ and in the matrix elements) and $E_{cv}=E_c-E_v$, and the shift vector~\cite{Sipe2000},
\begin{equation}
\sigma^{yyy}(\omega)=-\frac{e^{3}}{4\pi\hbar}\,g_{s}g_{v}\!
\int\! d^{2}k\; |r^{y}_{cv}|^{2}R^{y,y}(\vk)\,
\delta\!\left(E_{cv}-\hbar\omega\right),
\label{eq:absunits}
\end{equation}
where $R^{a,y}=\partial_{k_a}\arg r^{y}_{vc}+(A^{a}_{cc}-A^{a}_{vv})$ is
the displacement along $a$ of a carrier photoexcited by $y$-polarized light,
$A^{a}_{nn}=\langle u_n|i\partial_{k_a}u_n\rangle$ being the intraband Berry
connection (with $u_n$ the cell-periodic part of $\psi_n$); its sign follows the $(v,c)$ ordering of the windings, and
$\bm R=(R^{x,y},R^{y,y})$ is the shift-vector field of the $y$ dipole.
Following the quantization framework of intra- and inter-band Berry phases in
the shift current~\cite{Alexandradinata2024,Zhu2024}, on a disk $S_{\rho}$ of
radius $\rho$ centered at $K$ we use three quantities built from this field and
from the Berry curvatures $\Omega_n$ of the two bands:\ the optical vorticity
$\nu(\rho)$, the interband Berry flux $F(\rho)$, and their difference, the interband
character $\chi$,
\begin{equation}
\begin{split}
 \nu(\rho)=\frac{1}{2\pi}\oint_{\partial S_{\rho}}\!\bm{R}\cdot d\vk,&\quad
F(\rho)=\frac{1}{2\pi}\!\int_{S_{\rho}}\!(\Omega_c-\Omega_v),\\
\chi&=\nu-F.
\label{eq:character}
\end{split}
\end{equation}
Neither $\nu$ nor $F$ is quantized by itself; $\chi$ is an integer that counts
the vortex charges enclosed by the disk, as Sec.~\ref{sec:counting} shows, and
it is the quantity plotted in the staircases of Fig.~\ref{fig:summary}(b,d).

We evaluate Eq.~(\ref{eq:absunits}), and separate its channels, through
the sum-rule form of Sipe and Shkrebtii~\cite{Sipe2000},
$|r^{y}_{cv}|^{2}R^{y,y}\equiv-I_{yyy}$ with
\begin{align}
I_{yyy}(\vk) &= -\frac{1}{\omega_{cv}^{2}}\,
\mathrm{Im}\!\left[v^{y\,*}_{cv}\left(T_{\rm dir}+T_{\rm virt}\right)\right],
\label{eq:integrand}\\
T_{\rm dir} &= \frac{2 v^{y}_{cv}\,\Delta^{y}_{cv}}{\omega_{cv}}-v^{yy}_{cv},
\nonumber\\
T_{\rm virt} &= \sum_{p\neq c,v} v^{y}_{cp}v^{y}_{pv}
\left[\frac{1}{\omega_{pv}}-\frac{1}{\omega_{cp}}\right],
\nonumber
\end{align}
with $\Delta^{y}_{cv}=v^{y}_{cc}-v^{y}_{vv}$ and
$v^{yy}_{cv}=\langle c|\partial^{2}H/\partial k_y^{2}|v\rangle$.  The split into a two-band (\emph{direct}) and a
spectator-band (\emph{virtual}) channel is the diagnostic used
throughout; $I_{xyy}$, with the derivative taken along $x$, gives
$R^{x,y}$ in the same way.

In Eq.~(\ref{eq:absunits}), $g_{s}g_{v}=4$ ($g_s=g_v=2$ are the spin and valley degeneracies) and the Dirac delta $\delta(\cdot)$, is
replaced by a Gaussian of width $\eta=10$~meV unless stated otherwise.  The
integral runs over a disk of radius $0.15$~\AA$^{-1}$ around $K$; the $K'$
valley, which contributes identically by time reversal
[$H(-\vk)=H^{*}(\vk)$ gives $I_{yyy}(-\vk)=I_{yyy}(\vk)$] even though its
vortex charges are opposite, enters through $g_v$; the vorticity is valley
odd, the photocurrent valley even.  We quote $\sigma^{yyy}$ in $\mu$A\,\AA/V$^{2}$
($1~\mu$A\,\AA/V$^{2}=10^{-16}$~A\,m/V$^{2}$), with $\sigma^{yyy}>0$ denoting
dc current along $+\hat{y}$ for light polarized along $\hat{y}$.  All results
are for the clean, zero-temperature limit;
finite temperature enters only through the occupation factors and an extra
thermal contribution to $\eta$, neither of which affects the zero counting.

\section{The chiral expansion and the four rules}
\label{sec:rules}

\begin{figure*}
\includegraphics[width=\textwidth]{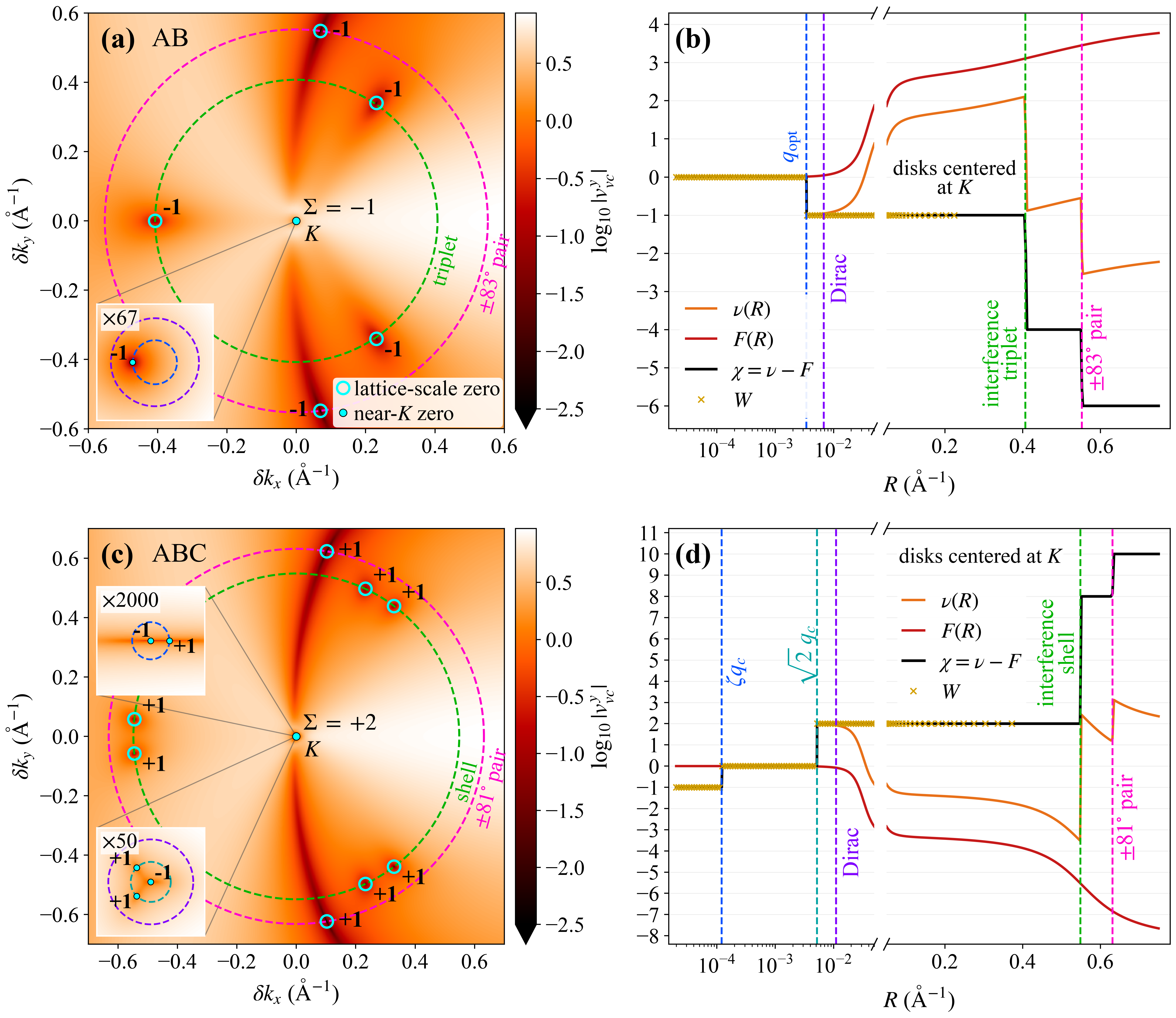}
\caption{Optical zeros and their quantized counting in the biased AB
bilayer ($\gamma_3=0.38$~eV, $\delta=0.15$~eV; top) and ABC trilayer
($\gamma_3=0.30$~eV, $U=0.05$~eV; bottom).  (a),(c)~$\log_{10}|v^{y}_{vc}|$
($v$ in eV\,\AA) around $K$ (axes $\delta\bm k=\bm q=\vk-K$), with the optical zeros of the $y$ dipole and their windings:\
open circles at lattice scale, filled dots near $K$.  Dashed circles mark the
constellation radii in the insets and the lattice-scale zeros in the main panels, drawn as vertical lines of the same color in (b),(d).
The lattice-scale zeros (Sec.~\ref{sec:counting} and Appendix~\ref{app:census}) form the
interference ``triplet'' in AB and a six-zero interference ``shell'' in ABC (green), plus one
additional pair in each case, the ``$\pm83^{\circ}$ pair'' in AB and the ``$\pm81^{\circ}$ pair''
in ABC (magenta).  The insets
(zoom factors indicated) resolve the near-$K$ zeros:\ in (a), the vortex at
$q_{\rm opt}$ [Eq.~(\ref{eq:qopt})] inside the satellite-Dirac radius $q_{\rm D}$ (labeled ``Dirac''); in (c), the protected
zero at $K$, with pinned charge $w_K=-1$ (Rule~4), the bias satellite at $\zeta\qc$, the two warping satellites
at $\sqrt2\,\qc$, and the satellite-Dirac radius $q_{\rm D}=3\qc$
[$\qc$ and $\zeta$ of Eqs.~(\ref{eq:qc}) and~(\ref{eq:zeta})].  In the main panels these zeros
are unresolved, and the dot at $K$ carries their total charge
$\Sigma=\sum_i w_i$ (Rule~1).
(b),(d)~The enclosed-charge staircase $\chi=\nu-F$ on disks of radius $\rho$
centered at $K$ [Eq.~(\ref{eq:character})], with $\nu$, $F$, and the direct
winding $W$ (crosses, shown only where its fixed-reference gauge is well
defined; Appendix~\ref{app:census}); the radial axis is logarithmic near $K$ and
linear at lattice scale (the disk radius $\rho$ is labeled $R$ in the figure axes).}
\label{fig:summary}
\end{figure*}

A single downfolded expansion of the $N$-layer Hamiltonian generates
the whole constellation of optical zeros.  We set that expansion down, read
four rules off it, and work the trilayer out in full; the biased bilayer and
trilayer of Fig.~\ref{fig:summary} are the examples we return to throughout.

\subsection{The L\"owdin path expansion}

The momentum measured from the valley can be written as the complex number
$\xi=q_x+iq_y$, with $\vk=K+\bm q$ and $q=|\bm q|$.  L\"owdin
downfolding~\cite{Lowdin1951,McCannKoshino2013} of the
$2N\times2N$ Hamiltonian onto $\{A_1,B_N\}$ organizes the
off-diagonal element $\bar h_N(\xi)$ as a sum over paths: each intralayer hop
contributes a factor $\xi$ (velocity $v_0=\tfrac{\sqrt3}{2}a\gamma_0$), each
skew hop a factor $\xi^{*}$ (velocity $v_3=\tfrac{\sqrt3}{2}a\gamma_3$), and each dimer a factor
$\gamma_1^{-1}$,
\begin{equation}
\bar h_N(\xi)=\sum_{s\ge0}(-1)^{N-s-1}\,C_s\,
\frac{v_0^{\,N-2s}v_3^{\,s}}{\gamma_1^{\,N-s-1}}\;
\xi^{\,N-2s}(\xi^{*})^{s},
\label{eq:hN}
\end{equation}
where $C_s=\binom{N-s}{s}$ counts skew placements ($C_1=N-1$;
Appendix~\ref{app:lowdin}; the alternating sign is that of Eq.~(\ref{eq:Ck})).  We write the chiral term as $\xi^{N}$; the
conjugate element, which carries $\xi^{*N}$, has all windings reversed, see Sec.~\ref{sec:models}.

The index $s$ counts the skew hops along a path.  The $s=0$ term is the
chirality term $\propto\xi^{N}$, $s=1$ is the leading warping correction, and
the sum terminates at the maximal-skew term $s=\lfloor N/2\rfloor$ (the
pure-skew path for even $N$).  Each skew hop trades a factor $\xi$ for
$\xi^{*}$ and one power of $v_0$ for $v_3$, so the ratio of successive terms,
$\gamma_1v_3/(v_0^{2}q)$, is the same for every $N$ and grows as the valley is
approached.  Two statements follow.  At finite $q$, the constellation is set
by the competition between the chirality term and the warping corrections.  At
$K$ itself, the monomial of lowest degree, the maximal-skew one, outgrows all
the others, and since the interband dipole is a momentum derivative of
$\bar h_N$, that single term decides whether the transition is dark and with
what winding.

\subsection{The four rules}
Each rule below is stated with the term of Eq.~(\ref{eq:hN}) that produces
it.

\emph{Rule 1 (charge).}  The total vortex winding near $K$, per linear
polarization, is $\sum_i w_i=\pm(N-1)$, with $w_i$ the charges of the individual zeros.  This is the winding of the chiral matrix
element that underlies the exciton selection rules of gapped chiral systems,
derived analytically for the isotropic chiral model in circular
polarization~\cite{Zhang2018} and found in ab initio matrix elements of
gapped graphene stacks with $N\le3$~\cite{Cao2018}.  The same
order-$(N-1)$ optical vortex at $\xi=0$ appears in the isotropic massive
$N$-Dirac model for any mass, including its realization in rhombohedral
graphene~\cite{Sukhachov2026}.  Here we recover it, consistently with these
works, for linear polarization and at finite $q$ from the chirality term of
Eq.~(\ref{eq:hN}), $\partial_{k}(\xi^{N})\sim\xi^{N-1}$.  What the path
expansion adds is that the warping corrections ($s\ge1$), which act only at
finite $q$, redistribute this charge into zeros at finite momentum without
changing its sum, since the winding on a loop enclosing the constellation is
an integer that changes only when a zero crosses the loop.  We verify this
conservation numerically, for each vortex, in every warped constellation for
$N=2$--$6$.  The overall sign is fixed by which low-energy sublattice hosts
the valence band, i.e.\ by the sign of the bias.  ``Near $K$'' means within
the constellation window; zeros at lattice-scale radii have a different
origin and are treated in Sec.~\ref{sec:counting}.

\emph{Rule 2 (protection).}  For $N\ge3$ the $K$ point remains an
\emph{exact} optical zero for arbitrary $\gamma_3$ and $U$, which we
confirm numerically. With no direct skew bridge between the low-energy sites (Sec.~\ref{sec:models}), every monomial of
Eq.~(\ref{eq:hN}) is at least quadratic in $q$ for $N\ge3$, so its momentum
derivative, and with it the dipole, vanishes at $K$; the bias enters only
through the L\"owdin denominators and cannot lower that degree.  Warping can
therefore redistribute vortex charge but never brighten $K$.  The Bernal
bilayer is the only exception:\ its skew hop is the direct bridge, contributes
a term linear in $\xi^{*}$, and expels the single vortex to
\begin{equation}
q_{\rm opt}=\frac{\gamma_1\gamma_3}{3 a_{cc}\gamma_0^{2}},
\qquad
q_{\rm D}=2\,q_{\rm opt},
\label{eq:qopt}
\end{equation}
on the $k_x$ axis (opposite sides for $x$ and $y$ polarization), a position
independent of $\delta$, with $q_{\rm D}$ the distance from $K$ to the satellite Dirac points.

\begin{figure}
\includegraphics[width=\columnwidth]{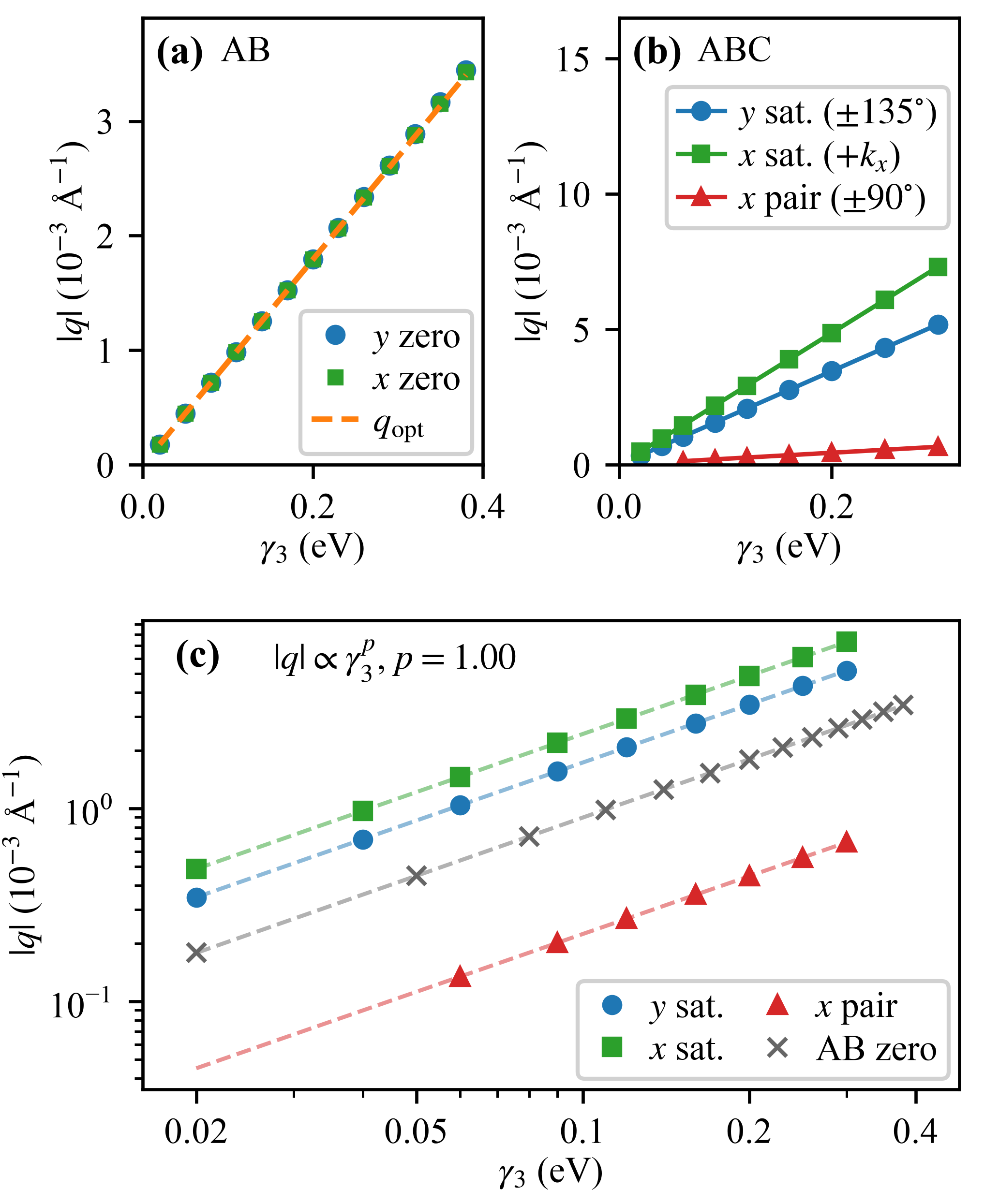}
\caption{Vortex and satellite positions versus $\gamma_3$ (AB:\
$\delta=0.15$~eV; ABC:\ $U=0.05$~eV).  (a)~The bilayer $y$- and
$x$-polarization zeros and the analytic law $q_{\rm opt}$
[Eq.~(\ref{eq:qopt})].  (b)~The trilayer $y$ satellites ($\pm135^{\circ}$),
the $x$ satellite on the $+k_x$ axis, and the bias-induced $x$ pair
($\pm90^{\circ}$).  (c)~The same data on log--log axes with power-law fits
(dashed); all fitted exponents equal $1.00$.  Momenta in units of $10^{-3}$~\AA$^{-1}$; $p$ in (c) is the fitted exponent of $|q|\propto\gamma_3^{p}$.}
\label{fig:traj}
\end{figure}

\emph{Rule 3 (scale).}  All satellite radii of the warped constellations are
linear in $\gamma_3$ and set by the single scale
$\gamma_1\gamma_3/(a_{cc}\gamma_0^{2})$, with $N$-dependent $O(1)$
coefficients.  Each satellite sits where the chirality term balances a
warping correction, and because the ratio of successive terms is the same
for every $N$, the balance radius collapses onto that
one combination; the trilayer case is worked out in Sec.~\ref{sec:trilayer}.
Figure~\ref{fig:traj} follows the positions across the family as $\gamma_3$
is varied and confirms the linear law.

\emph{Rule 4 (pinned charge).}  The charge left pinned at $K$ in the warped
constellation is fixed by the maximal-skew term of the path expansion:
$w_K=-(N-2)/2$ for even $N$ and $w_K=-(N-5)/2$ for odd $N\ge5$, both
independent of the bias.  The bias cannot change $w_K$ at any $N\ge4$,
for the same reason that it cannot brighten $K$:\ it cannot lower the minimal
monomial degree (it does redistribute satellite charge for odd $N\ge5$).  The trilayer is the unique
\emph{marginal} member
of the family: its leading term is charge-degenerate, and the bias resolves
it to $w_K=-1$ for any $U\neq0$, with bias-induced satellites carrying the
difference (Sec.~\ref{sec:fine}).  We confirm the rule numerically for $N$ up to $9$ (Appendix~\ref{app:oddN},
which also shows that the even and odd branches form a single arithmetic
ladder).

The monolayer ($N=1$) is the limiting case:\ Rule~1 assigns it zero net
charge, and its valley is in fact empty.  Its chirality and warping terms both
descend from the intralayer hopping and share the single scale $\gamma_0$, so
their balance falls at lattice-scale momentum, outside the expansion, where
the full model leaves no zero (Appendix~\ref{app:lowdin}).  Near-$K$ optical
zeros thus require competing terms with \emph{independent} energy scales,
which the interlayer couplings $\gamma_1,\gamma_3,\gamma_2,U$ supply; the
hierarchy $\gamma_1\gamma_3/\gamma_0^{2}\ll1$ places the resulting vortices at
small momentum, optically accessible at the gap edge.

\subsection{Trilayer constellation}
\label{sec:trilayer}

The trilayer is the smallest member in which a protected $K$ zero, the
warping satellites, and the bias fine structure coexist, so we work its
constellation out in full.  For $N=3$,
\begin{equation}
\bar h_3(\xi)=\frac{v_0^{3}}{\gamma_1^{2}}\,\xi^{3}
-\frac{2v_0v_3}{\gamma_1}\,|\xi|^{2}
\equiv c_3\xi^{3}-b\,|\xi|^{2}.
\label{eq:h3}
\end{equation}
The warping term is \emph{non-chiral} ($|\xi|^{2}$ carries zero winding),
in sharp contrast to the bilayer.  Setting
$v^{y}=\partial_{k_y}\bar h_3=3ic_3\xi^{2}-2b\,q_y$ and
$v^{x}=3c_3\xi^{2}-2b\,q_x$ to zero gives, with
\begin{equation}
\qc=\frac{b}{3c_3}=\frac{2v_3\gamma_1}{3v_0^{2}}
=\frac{4\gamma_1\gamma_3}{9\,a_{cc}\gamma_0^{2}},
\label{eq:qc}
\end{equation}
(i) two $y$-polarization satellites at $(-\qc,\pm \qc)$, i.e.\ radius
$\sqrt2\,\qc$ at $\pm135^{\circ}$; (ii) one $x$ satellite at $(+2\qc,0)$; and
(iii) satellite Dirac points ($\bar h_3=0$) at radius $q_{\rm D}=3\qc$ and angles
$0,\pm120^{\circ}$.  The optical zeros, at $\sqrt2\,\qc$ and $2\qc$, and the
satellite Dirac points of the $U=0$ spectrum, at $3\qc$, are different objects
(dark transitions with an open gap, and gap closings), but they are features of
the single function $\bar h_3$ and scale with the single momentum $\qc$; their
radii stand in the ratios $\sqrt2:2:3$, in quantitative agreement with the full
TB calculations [Fig.~\ref{fig:traj}(b)].

\subsection{Bias-induced fine structure of the trilayer}
\label{sec:fine}

At $U=0$ the trilayer constellation is only partly resolved.  Writing
$q_y=(\xi-\xi^{*})/2i$, the part of $v^{y}$ linear in momentum is
$ib\,(\xi-\xi^{*})$, the sum of a holomorphic piece, $ib\,\xi$, whose phase
winds once counterclockwise around $K$ (winding $+1$), and an antiholomorphic
one, $-ib\,\xi^{*}$, winding once clockwise (winding $-1$), of exactly equal
modulus.  Neither
dominates, so that $v^{y}$ is real along $q_y=0$ and the zero at $K$ is not
isolated at this order; the winding at $K$ is therefore undetermined, and this
is the charge degeneracy that makes the trilayer marginal in Rule~4.  Resolving it requires
the L\"owdin denominators evaluated band by band at $E_{c,v}=\pm U$; the
corrections linear in $U$ cancel exactly (Appendix~\ref{app:lowdin}), and the expansion parameter is
\begin{equation}
\zeta=\frac{2U^{2}}{\gamma_1^{2}}.
\label{eq:zeta}
\end{equation}
To linear order in $\xi$,
\begin{equation}
v^{y/x}_{vc}\;\propto\;3c_3\xi^{2}\;\pm\;b\,\xi\;-\;b(1+\zeta)\,\xi^{*}.
\label{eq:fine}
\end{equation}
Three consequences follow.  (i)~The bias lifts the degeneracy:\ the linear
part of Eq.~(\ref{eq:fine}), $\alpha\xi+\beta\xi^{*}$ with $\alpha=\pm b$ and $\beta=-b(1+\zeta)$, has
$|\beta|>|\alpha|$ for any $U\neq0$, so the antiholomorphic component
dominates and $w_K=-1$ for both polarizations.  (ii)~A $y$ satellite
detaches along $+k_x$ at $q=\zeta\,\qc\propto U^{2}$.  (iii)~An $x$ pair
detaches at $\big({-\tfrac{\zeta}{2}\qc},\,\pm \qc\sqrt{\zeta(2+\zeta)/2}\big)$,
radius $\simeq\sqrt2\,\qc\,U/\gamma_1\propto U$, at angles
$90^{\circ}{+}\vartheta$ with $\tan\vartheta\simeq U/(\sqrt2\gamma_1)$.  The full TB calculations confirm the positions, the scalings, and the angle
drift, and the
charges add up
exactly: $-1+1+2=+2$ ($y$) and $-1+2+1=+2$ ($x$).  At $\gamma_3=0$
[$b=0$ in Eq.~(\ref{eq:fine})] no fine structure exists at any $U$: the
phenomenon is a warping--bias interference.

\section{Quantized counting: the character staircase} 
\label{sec:counting}

We count the near-$K$ zeros with the character of Eq.~(\ref{eq:character}),
evaluated on disks $S_{\rho}$ of growing radius.  Integrating the circulation
identity of Ref.~\cite{Zhu2024} [Eq.~(17) there] over $S_{\rho}$ shows why
$\chi$, and not $\nu$, is quantized.  The shift vector is a phase gradient
minus an intraband-connection mismatch, so
\begin{equation}
\begin{split}
\nu(\rho)&=\underbrace{\frac1{2\pi}\oint_{\partial S_{\rho}}\!\nabla\arg r^{y}_{vc}
\cdot d\vk}_{\textstyle \sum_i w_i}\\
&\quad+\;\underbrace{\frac1{2\pi}\oint_{\partial S_{\rho}}\!
(\bm A_{cc}-\bm A_{vv})\cdot d\vk}_{\textstyle F(\rho)} .
\end{split}
\label{eq:chisplit}
\end{equation}
The first term is the circulation of the gradient of a phase, which counts the
windings of every dipole zero enclosed, exactly the charges of
Sec.~\ref{sec:rules}.  The second is the circulation of the interband
connection difference, which by Stokes' theorem is the enclosed interband
Berry flux $F$.  Hence $\nu$ mixes the two and is \emph{not} quantized by
itself, whereas
\begin{equation}
\chi=\nu-F=\sum_i w_i
\end{equation}
is:\ the flux cancels and the character returns the enclosed vortex charge as
an integer~\cite{Alexandradinata2024}.  This is what makes $\chi(\rho)$ a Gauss
law for the constellation:\ it can change only when the growing disk encloses
a new zero, and it changes by exactly that zero's charge.

Figure~\ref{fig:summary} shows the four rules acting together and the
staircases that count them.  In the trilayer [panels (c),(d)], $K$ stays dark
(Rule~2) with pinned charge $w_K=-1$ (Rule~4); $\chi$ starts at this value,
steps to $0$ at the bias satellite $\zeta\qc$ (blue dashed circle), more than
an order of magnitude below the warping scale, and reaches $+2=N-1$ (Rule~1)
at the two warping satellites at $\pm135^{\circ}$ and radius $\sqrt2\,\qc$
(teal dashed circle; Rule~3).  In the bilayer [panels (a),(b)], the exception
to Rule~2, nothing is pinned at $K$, and the single step $0\to-1$, negative by
the convention of Sec.~\ref{sec:models}, falls at $q_{\rm opt}$.  Every
plateau is integer to numerical precision and every step sits at an analytic
radius.  The direct winding $W$ (crosses) agrees with $\chi$ wherever it is
defined.  We verified in the same way the tetralayer staircase, which climbs
$-1\to0\to+1\to+3$ as Rules~1 and~4 require, and the $\gamma_3=0$ bilayer,
for which $\chi\equiv-1$ at all radii (not shown).

Although $\nu$ is not quantized, it \emph{appears} quantized out to $\rho\sim10^{-2}$~\AA$^{-1}$ in
Fig.~\ref{fig:summary}(b,d).  The reason is that the chirality that flattens
the bands also flattens their quantum geometry:\
$\Omega_c-\Omega_v\propto q^{2N-2}$, so $F(\rho)\sim\rho^{2N}$ stays negligible
until the disk reaches the band-edge ring where the curvature
sits, at $q_{\rm edge}\sim(U\gamma_1^{2}/v_0^{3})^{1/3}\approx0.03$~\AA$^{-1}$
for the trilayer ($0.016$--$0.026$~\AA$^{-1}$ around the warped band-edge ring of
the full model). 

The constellation thus lives in a geometrically flat window
that holds all of the charge and none of the flux, which lets $\nu$ pass for a
quantum number in small-disk or effective two-band treatments; only $\chi$
survives the crossing of the band-edge ring.  The two scales are independently
tunable ($\qc\propto\gamma_3$, $q_{\rm edge}\propto U^{1/3}$), so the window
can be closed at will, and a two-band calculation indicates that the staircase
is model independent even though $\nu$ and $F$ separately are not
(Appendix~\ref{app:census}, Fig.~\ref{fig:twoband}).  

Beyond the band-edge ring the
flux saturates to the interband valley Chern flux $F\to\pm N$, with the
spectator pairs adding a smooth nonquantized background.  Band topology (Chern
numbers, which cannot resolve $\gamma_3$) and optical topology ($\chi$, reshaped by every
constellation parameter) are therefore cleanly separated, an explicit
realization across a material family of the distinction emphasized in
Ref.~\cite{Alexandradinata2024}.

Beyond the constellation, the staircase
acquires further steps, on the linear axis of Fig.~\ref{fig:summary}(b,d),
whose origin is not chirality. Away from $K$ the dimer and skew bonds couple
the layers with different momentum factors, and optical zeros appear where
the two channels interfere destructively.  This happens on the ring
$|f|=\gamma_1/\gamma_3$ at three discrete angles per valley, which in the
bilayer produces an interference triplet (Appendix~\ref{app:census}).  Two
properties set these zeros apart from the constellation.  They tune the other
way, moving \emph{inward} as the warping grows while $\qc\propto\gamma_3$
moves outward; and they have a threshold, since $|f|\le3$ leaves no solution
below $\gamma_3^{*}=\gamma_1/3\approx0.13$~eV, whereas the constellation
exists for arbitrarily small $\gamma_3$.  These shells (Fig.~\ref{fig:summary}) take the trilayer staircase
across more than three orders of magnitude in radius and close the global
charge balance,
$\sum_i w_i+w_\Gamma=0$ per polarization, with the sum running over the zeros of both valleys and $w_\Gamma$ the winding at the $\Gamma$ point, as time reversal requires; their
positions, charges, and census are collected in Appendix~\ref{app:census}.

\section{Remote hoppings and the gate-tunable merger} 
\label{sec:Nlayer}

The rules were derived with the minimal couplings $\gamma_0$,
$\gamma_1$, and $\gamma_3$.  The remote couplings of Sec.~\ref{sec:models}
renormalize their scales without moving charge, with one exception:\
$\gamma_2$, whose competition with the bias drives a gate-tunable vortex
merger.

\subsection{Robustness against the remote hoppings}

The couplings $\gamma_4$ and $\Delta'$ enter only at second order
(Appendix~\ref{app:lowdin}).  Numerically, the bilayer vortex moves by a
fraction $-(\gamma_4/\gamma_0)^{2}$ to three digits, with no linear term and
no dependence on $\delta$, while in the trilayer $\gamma_4$ and $\Delta'$
shift all constellation radii by the common fraction $-2(\gamma_4/\gamma_0)^{2}$.  They renormalize the
single scale $\qc$, so the radii move but their ratios $\sqrt2:2:3$ do
not.

The coupling $\gamma_2$ requires a separate argument, because in the
trilayer it is the only direct bond between the two low-energy sites.  It is
a vertical hop and hence momentum independent:\ it shifts $\bar h$ by a
constant but is invisible to $\partial H/\partial k$, so it cannot brighten
$K$.  Even the \emph{unbiased} trilayer, gapped by $\gamma_2$ alone
[$E_g(K)=|\gamma_2|$], keeps its dark transition at $K$.  With the full
parameters fitted to density-functional theory (DFT) in Ref.~\cite{Zhang2010} ($\gamma_1=0.502$,
$\gamma_3=-0.377$, $\gamma_4=-0.099$, $\gamma_2=-0.0171$, and
$\Delta'=0.0014$~eV, minus the on-site parameter of that reference), the $K$ point remains an exact
optical zero with $w_K=-1$ and total charge $+2$; the warping satellites stay
at $\sqrt2\,\qc$, rotated by $180^{\circ}$ by the sign of $\gamma_3$, and only
the bias satellite moves, through the eigenvector mixing that $\gamma_2$
induces (Sec.~\ref{sec:merger}).  For $N\ge5$ the statement is weaker:\ there
$\gamma_2$ couples the low-energy sites to the dimer manifold
($A_l$--$B_{l+2}$) and weakly brightens $K$ [$|v^{y}(K)|\propto|\gamma_2|$],
but it creates no near-$K$ vortex and leaves the pinned winding and the total
charge intact (checked for $N=5$ and $7$; Appendix~\ref{app:oddN}).

\subsection{The gate-tunable vortex merger}
\label{sec:merger}

\begin{figure}
\includegraphics[width=\columnwidth]{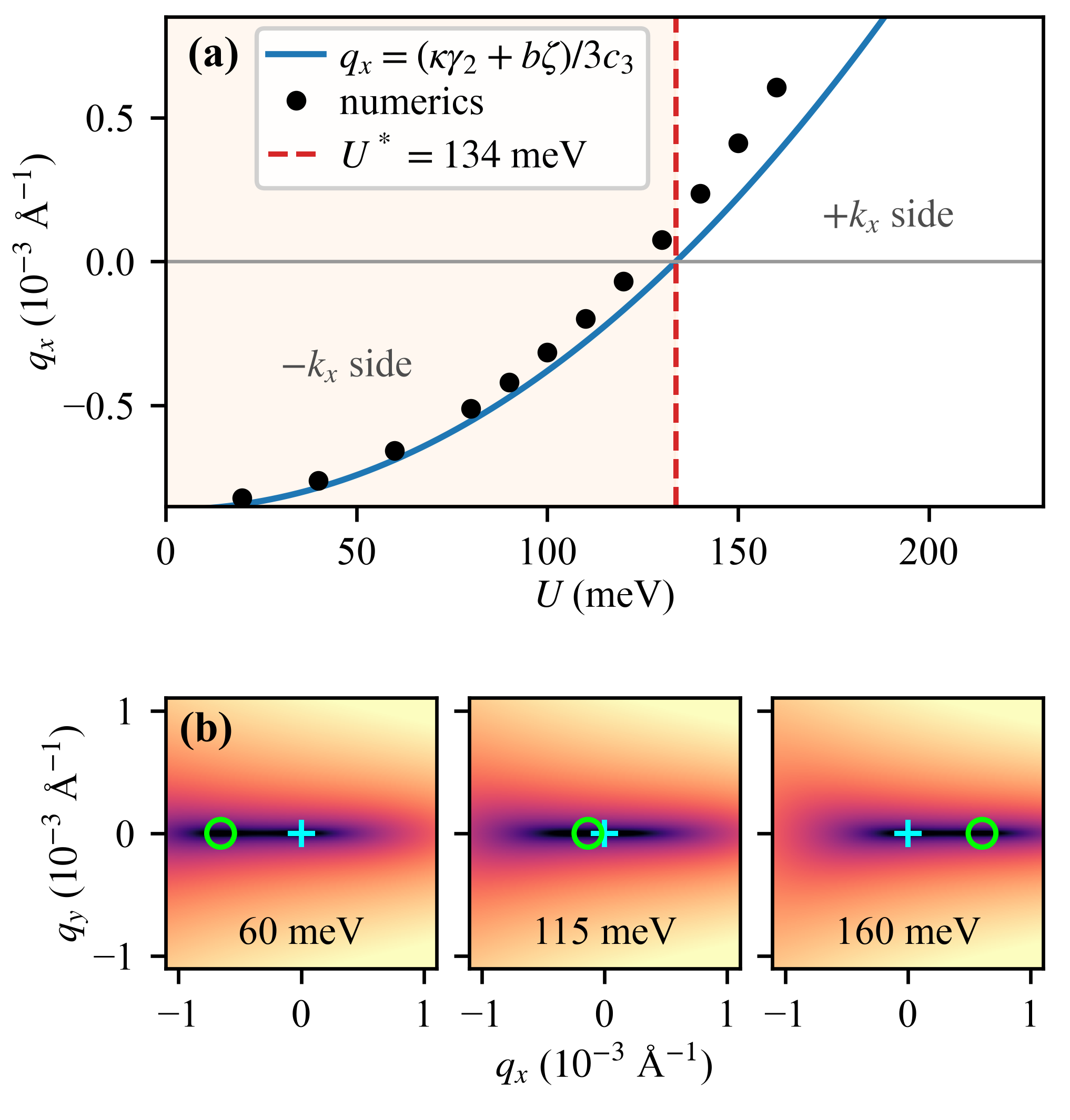}
\caption{Electrically tunable vortex merger in the ABC trilayer
($\gamma_2=-17.1$~meV; other parameters as in Sec.~\ref{sec:models}).
(a)~Signed on-axis position of the bias satellite
versus bias $U$: numerics (points) and the leading-order law
$q_x=(\kappa\gamma_2+b\zeta)/(3c_3)$ (solid); the dashed line marks the leading-order $U^{*}$.
(b)~Maps of $\log_{10}|v^{y}_{vc}|$ at $U=60$, $115$, and $160$~meV,
with $K$ (cross) and the bias satellite (circle); $v$ in eV\,\AA.  Momenta $\bm q=\vk-K$ in units of $10^{-3}$~\AA$^{-1}$.}
\label{fig:merger}
\end{figure}

Through the eigenvector mixing induced by
$\bar h(K)=\gamma_2/2$, the on-axis linear coefficient of Eq.~(\ref{eq:fine})
becomes $\alpha+\beta=-(\kappa\gamma_2+b\,\zeta)$, with
$\kappa=(v_0^{2}+2v_3^{2})/\gamma_1^{2}$, so the bias satellite sits at
$|\kappa\gamma_2+b\zeta|/(3c_3)$.  The law holds numerically, including for
the full DFT set, where $\gamma_4$ and $\Delta'$ enter the effective model only
through terms proportional to the identity and drop out.  Since $\gamma_2<0$
while $b\zeta>0$, the two terms compete, and at
\begin{equation}
U^{*}=\gamma_1\sqrt{\kappa|\gamma_2|/(2b)}\;\approx\;0.13~\text{eV}
\end{equation}
the bias satellite is driven \emph{through} the protected $K$ vortex,
momentarily neutralizing its charge and re-emerging on the opposite side
(Fig.~\ref{fig:merger}).  This electrically tunable vortex merger appears as
the collapse of the first step of the $\chi(\rho)$ staircase of
Sec.~\ref{sec:counting}. The merger takes place within ${\sim}10^{-3}$~\AA$^{-1}$ of $K$ and
leaves no measurable trace in $\sigma^{yyy}(\omega)$, to which that region
contributes less than $5\times10^{-4}$ of the band-edge peak:\ it is a property
of the constellation, not of the photocurrent.

\section{Imprint on the shift current}
\label{sec:shift}

\subsection{\texorpdfstring{The vortex imprint on the integrand}{The vortex imprint on the integrand}}
\label{sec:imprint}

\begin{figure}
\includegraphics[width=\columnwidth]{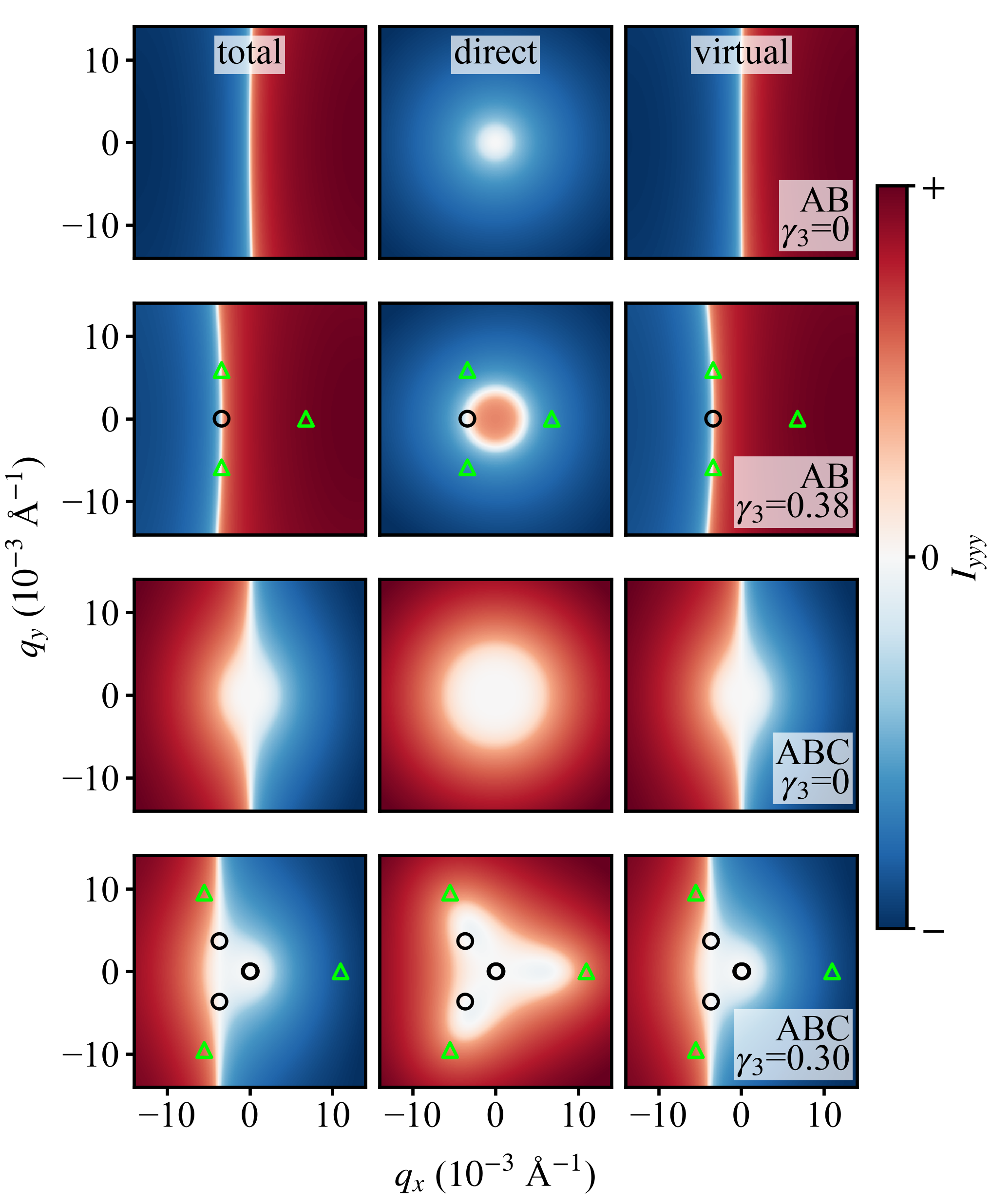}
\caption{Shift-current integrand $I_{yyy}(\vk)$ near $K$ over a
$\pm0.014$~\AA$^{-1}$ window, on a symmetric-logarithmic color scale
normalized separately in each panel (red and blue denote sign).  Rows:\ AB
($\delta=0.15$~eV) and ABC ($U=0.05$~eV), each without ($\gamma_3=0$) and with
warping ($\gamma_3=0.38$~eV for AB and $0.30$~eV for ABC).
Columns:\ total, direct, and virtual channels.  In the warped rows, circles
mark the $y$-polarization optical zeros and green triangles the satellite
Dirac points.  Axes $q_x$ and $q_y$, with $\bm q=\vk-K$, in units of $10^{-3}$~\AA$^{-1}$.}
\label{fig:integrand}
\end{figure}

The band-edge response of these stacks is dominated by the virtual,
multiband channel~\cite{Chen2024}, which trigonal warping switches on,
amplifying the peak by an order of magnitude and setting its
sign~\cite{SelectionRule2026}.  Figure~\ref{fig:integrand} shows what the
constellation does to it.  The nodal structure of
$I_{yyy}$ is anchored on the optical zeros, where $I_{yyy}=-|r|^{2}R$ vanishes
while $R$ diverges, so the constellation is visible in the integrand itself.
It is not visible in $\sigma^{yyy}(\omega)$, which is a smooth curve:\ the
$|r|^{2}$ weighting of Eq.~(\ref{eq:absunits}) removes every divergence.

In every row the total integrand reproduces the sign
pattern of the virtual channel rather than that of the direct one.  Warping
displaces the nodal region of the total integrand together with the
zeros:\ in AB it passes through the vortex expelled to $q_{\rm opt}$ on
the $-k_x$ axis, and in ABC it encloses the protected $K$ zero and the two
warping satellites at $\pm135^{\circ}$.

\subsection{The linear law $|\sigma|_{\rm pk}\propto N$}
\label{sec:Nlaw}

\begin{figure}
\includegraphics[width=\columnwidth]{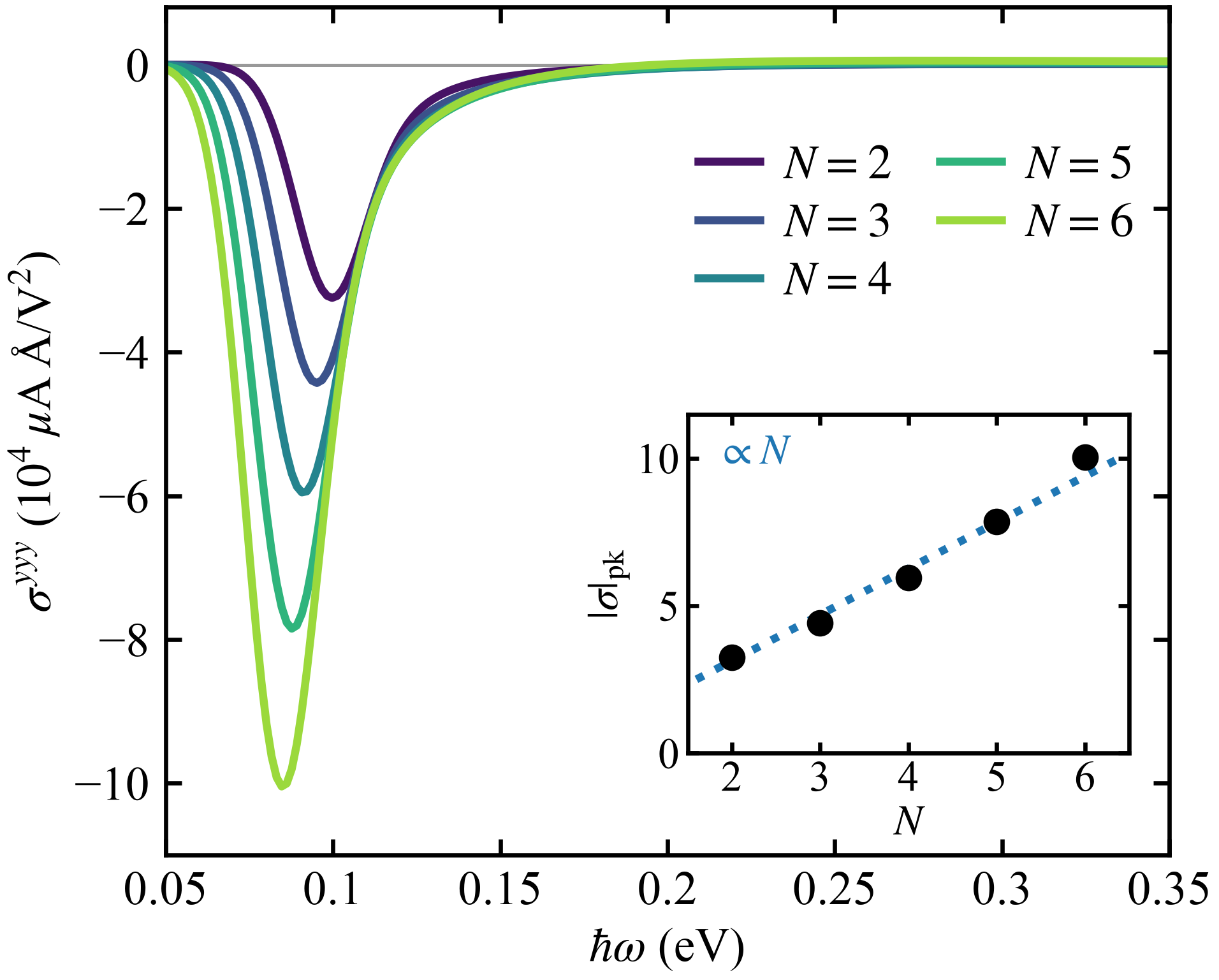}
\caption{Shift-current scaling across the rhombohedral ladder (full $2N\times2N$ tight-binding model, transition between the two low-energy bands, with the virtual sum of Eq.~(\ref{eq:integrand}) running over the $2N-2$ dimer bands; $U=0.05$~eV,
the $N$-layer hoppings of Sec.~\ref{sec:models} ($\gamma_3=0.30$~eV) for every $N$, including $N=2$, $\eta=10$~meV, absolute units; for $N=2$ the bias follows the ladder $\pm U$, not the $\delta$ convention of Sec.~\ref{sec:models}).
Main panel:\ $\sigma^{yyy}(\omega)$ near the band edge for $N=2$--$6$.
Inset:\ band-edge peak $|\sigma|_{\rm pk}$ versus layer number $N$, in the
same units, with the linear law $|\sigma|_{\rm pk}\simeq1.6\times10^{4}\,N$~$\mu$A\,\AA/V$^{2}$
(dotted).}
\label{fig:Nscaling}
\end{figure}

Figure~\ref{fig:Nscaling} extends the calculation to $N=2$--$6$.
The band-edge peak obeys $|\sigma|_{\rm pk}\propto N$, with a stable
virtual/direct ratio ${\simeq}15$ at the peak.  The $\gamma_3=0$ baseline (not shown)
grows faster (${\sim}{\times}2$ per layer), so the warping amplification falls
from ${\approx}{\times}45$ at $N=2$--$3$ to ${\times}9$ at $N=6$, while the
absolute response always increases.

The linear law has a derivable core (Appendix~\ref{app:Nlaw}).  At the
peak frequency every transition originates on the band-edge
ring of radius $q_{\rm edge}$, so the peak factorizes into
ring quantities, $|\sigma|_{\rm pk}\propto\langle I\rangle\,\mathcal{J}$, with
$\mathcal{J}(\omega)=\int d^{2}k\,g_{\eta}(\omega-\omega_{cv})$ the broadened
joint density of states of the ring ($g_{\eta}$ being the Gaussian of
Sec.~\ref{sec:models}) and $\langle I\rangle$ the
ring-averaged integrand.  Since the integrand, Eq.~(\ref{eq:integrand}),
carries a factor $1/\omega_{cv}^{2}$ and is dominated at the peak by the
virtual channel,
$\langle I\rangle\simeq\langle|v_{cv}|\rangle\,\langle|T_{\rm virt}|\rangle\,
|\cos\varphi|/\omega_{\rm pk}^{2}$, with $\omega_{\rm pk}$ the peak frequency and $\varphi$ the relative phase of the
two amplitudes.  Three statements
then fix the $N$-dependence.
(i)~\emph{The interband dipole on the ring carries one factor of $N$.}
The ring sits where the chiral coupling balances the bias,
$|\bar h_N(q_{\rm edge})|\simeq U$, and a degree-$N$ monomial is
radially steep, $|\partial_q\bar h_N|=N|\bar h_N|/q$; hence
$\langle|v_{cv}|\rangle\simeq c_1\,NU/q_{\rm edge}$ with $c_1$ of order
one:\ the dipole exceeds the geometric scale $U/q_{\rm edge}$ by precisely
the chirality index.
(ii)~\emph{The ring radius drops out.}  The band edge is a
one-dimensional van Hove ring of length $2\pi q_{\rm edge}$, so
$\mathcal{J}\propto q_{\rm edge}/\sqrt{\omega''\eta}$ ($\omega''$ the radial
curvature of the edge); its $q_{\rm edge}$ cancels the
$1/q_{\rm edge}$ of the dipole, and no geometric scale survives in the
product $\langle I\rangle \mathcal{J}$.
(iii)~\emph{The virtual amplitude sets the magnitude, not the trend.}
It is pinned at the dimer scale $v_0^{2}/\gamma_1$ at every $N$ and
contributes no factor of $N$ (Appendix~\ref{app:Nlaw}).
What remains is
\begin{equation}
|\sigma|_{\rm pk}\;\propto\;
N\,\frac{U\,\langle|T_{\rm virt}|\rangle\,|\cos\varphi|}
{\omega_{\rm pk}^{2}\,\sqrt{\omega''\eta}},
\label{eq:NlawMain}
\end{equation}
in which the single factor of $N$ is the dipole's: the modulus of the
same monomial $\bar h_N\propto\xi^{N}$ whose phase fixes the vortex charge
$N-1$.  The residual factors drift weakly and in mutually compensating
directions; this compensation is measured rather than derived, and it holds
$|\sigma|_{\rm pk}/N$ constant to $\pm7\%$.  At every bias the response
grows monotonically and at order
$N$, although the fitted exponent drifts between $0.8$ and $1.2$ as $U$ runs
from $0.03$ to $0.10$~eV, the edge-geometry factors of
Eq.~(\ref{eq:NlawMain}) shifting with the band-edge ring (Appendix~\ref{app:Nlaw});
the displayed law is the $U=0.05$~eV case.

The band-edge peaks themselves are broadening-limited.  They grow as
$\eta^{-0.7}$ over $\eta=2.5$--$40$~meV (AB at $\delta=0.15$~eV, ABC at
$U=0.05$~eV), neither the $1/\eta$ of an isolated
resonance nor the $\eta^{-1/2}$ of an ideal one-dimensional van Hove edge, a
signature of the anisotropic band-edge ring.  The frequency-integrated band-edge
weight, taken from $20$~meV below the band edge to $30$~meV above the
$K$-point transition energy, by contrast converges to within $2\%$ once
$\eta\lesssim10$~meV [$+61$ (AB) and $-1.39\times10^{3}$ (ABC)
eV\,$\mu$A\,\AA/V$^{2}$], and grows along the ladder as $N^{1.10}$ at
$\eta=5$ and $10$~meV, so the linear law is not an artifact of the
broadening.  In the clean limit at $\eta=10$~meV the peaks
(${\sim}10^{-12}$~A\,m/V$^{2}$) exceed by an order of magnitude the values
reported for unbiased Bernal (ABA) trilayers~\cite{FewLayer2026} and for prototypical
2D ferroelectrics~\cite{ChangSnS2023,NakamuraShift2024}, the combined effect
of the bias-induced band-edge van Hove singularity and the
$\gamma_3$-activated virtual channel~\cite{SelectionRule2026}.  For the
biased ABC trilayer, a six-band calculation with a fixed $\gamma_3$, a
$10$~meV Lorentzian broadening, and $U=0.2$--$0.4$~eV in our convention
reports band-edge peaks of about $5\times10^{-13}$~A\,m/V$^{2}$ at
$\hbar\omega\approx0.24$~eV~\cite{RTG2024}.  Our band-edge peak at
$U=0.05$~eV is about eight times larger; a factor of about six follows from
the $1/\omega_{cv}^{2}$ factor of Eq.~(\ref{eq:integrand}) for the smaller
gap ($\hbar\omega_{\rm pk}=0.096$ versus ${\approx}0.24$~eV), and the
remainder is within the differences in line shape and hopping parameters.
Because peak
heights depend on the sample as much as on the stack, comparisons across
samples and materials should use the integrated weight; the experimental
signatures of the law will be presented elsewhere.

\section{Discussion and outlook}
\label{sec:discussion}

The rhombohedral graphene family realizes a topological ladder for
resonant optical geometry:\ four analytic rules fix its constellation of
optical zeros, the character staircase counts it in integers, the
shift-current integrand carries it pointwise, and the band-edge shift current
grows linearly with $N$.

Three extensions suggest themselves.  First, excitonic corrections at
the band-edge ring, which will dress the band-edge response that a quantitative
comparison with mid-infrared BPVE experiments must target.  Second, strain,
which enters the effective theory as a valley-odd gauge field:\ it should
displace the constellations without breaking the protection rule, adding a
mechanical tuning parameter complementary to the bias.  Third, interactions in
the flat-band regime of larger $N$, where rhombohedral stacks develop
correlated phases:\ whether the symmetry-breaking orders they host can switch
the pinned charge, as the bias does in the trilayer.

The four rules are statements about the downfolded expansion of
Eq.~(\ref{eq:hN}) at leading order in $q$, verified numerically on the full
$2N\times2N$ lattice models of chirally stacked graphene gapped by a
perpendicular field; whether the same counting organizes other chiral
multiband systems is a question this work poses rather than settles.  The
radii at which the staircase steps fall are model quantities.  Near $K$ they
are robust, protected against $\gamma_4$ and $\Delta'$ to second order and
against $\gamma_2$ except for the pointwise darkness of $K$ at $N\ge5$; at
lattice scale the shell \emph{integers} are winding protected, while the
shell \emph{radii} are nearest-neighbor numbers.  The theory rests on a
single-particle Hamiltonian and a clean crystal, so disorder and many-body
corrections will dress the numbers quoted here:\ peak heights, the precise
radii, the location of the merger.  The counting is a different kind of
statement.  Vortex charges are integers fixed by the winding of a dipole
field, and the character $\chi=\nu-F$ that collects them changes only when a
zero crosses the contour; no smooth renormalization of the band structure can
move an integer.  What such corrections can do is move the zeros, and at most
create or annihilate them in neutral pairs; they cannot change the charge each
zero carries or the total the charges add up to.  In that sense the central
result stands:\ the optical geometry of a chirally stacked multilayer is
organized by one integer, which the constellation displays as a phase and the
photocurrent as an amplitude.

\begin{acknowledgments}
E.S.M. acknowledges financial support from Proyecto Interno USM 2026
PI\_LIR\_26\_12.

\emph{Data availability.} The data and codes that support the findings of this article are available from the authors upon reasonable request.

\emph{Use of AI tools.} During the preparation of this work, the authors used Claude (Anthropic; models Fable 5.1 and Opus 5.5) to write and run plotting scripts on the authors' workstation under their supervision, to check analytical steps against numerical evaluations and to edit the language of the manuscript. The authors verified every AI-assisted script and result, reviewed and edited all AI-assisted content, and take full responsibility for the content of this publication. No figure was generated or modified by AI image tools; all figures are plots of computed data.
\end{acknowledgments}

\appendix

\section{L\"owdin path expansion}
\label{app:lowdin}

This appendix derives the combinatorics of Eq.~(\ref{eq:hN}), the bias
parameter $\zeta$ of Eq.~(\ref{eq:fine}), and the second-order character of
the remote couplings (Sec.~\ref{sec:Nlayer}), and records why the monolayer
valley is empty.

Downfolding onto $P=\{A_1,B_N\}$ (with $Q$ the complementary set of dimer sites) gives the exact
effective Hamiltonian
$H_{\rm eff}(E)=H_{PP}+H_{PQ}(E-H_{QQ})^{-1}H_{QP}$~\cite{Lowdin1951}, in
which each bond carries its leading structure factor,
$-\gamma_0f\to v_0\xi^{(*)}$ and $-\gamma_3f^{*}\to v_3\xi^{(*)}$, with
$v_{0,3}$ as in Sec.~\ref{sec:rules}; the conjugation depends on the direction of the hop.  A path from $A_1$ to $B_N$ alternates
$A\!\to\!B$ moves (an intralayer hop of weight $v_0$, or a skew hop of weight
$v_3$ that also advances one layer) with dimer traversals $B_l\to A_{l+1}$,
each contributing $-1/\gamma_1$ at midgap.  A path with $s$ skew hops
therefore has $N-2s$ intralayer hops and $N-s-1$ dimer traversals, and any
$s$ of its $N-s$ $A\!\to\!B$ moves may be the skew ones, so
\begin{equation}
C_s=\binom{N-s}{s},\qquad
\text{sign}\propto(-1)^{N-s-1}.
\label{eq:Ck}
\end{equation}
Successive terms therefore alternate in sign [cf.\ Eq.~(\ref{eq:h3})],
and $C_{N/2}=1$ for the terminating pure-skew path at even
$N$.

Equation~(\ref{eq:fine}) follows once
the L\"owdin denominators are evaluated band by band.  For the biased trilayer
the two dimer blocks $\{B_1,A_2\}$ and $\{B_2,A_3\}$ (on-site energies $U,0$
and $0,-U$) have determinants $D_1(E)=(E-U)E-\gamma_1^{2}$ and
$D_2(E)=E(E+U)-\gamma_1^{2}$, which at the $K$-point band energies
$E_{c,v}=\pm U$ give
\begin{equation}
\begin{gathered}
D_1(E_c)=D_2(E_v)=-\gamma_1^{2}\ \ \text{(exact)},\\
D_1(E_v)=D_2(E_c)=-\gamma_1^{2}(1-\zeta),
\end{gathered}
\end{equation}
with $\zeta=2U^{2}/\gamma_1^{2}$.  The bias enters only through the
crossed evaluations, every correction linear in $U$ cancelling identically,
which is why $\zeta$ is quadratic in the bias.  The two $s=1$ paths of
$\bar h_3$ traverse different dimers, so exactly one of them is dressed by
$(1-\zeta)^{-1}\simeq1+\zeta$ in each band; expanding the resulting matrix
element to linear order in $\xi$ gives Eq.~(\ref{eq:fine}).

The same downfolding shows why $\gamma_4$ and
$\Delta'$ leave the constellation intact.  In the bilayer, the dimer block
$\{B_1,A_2\}$ has on-site energies $\varepsilon_B,\varepsilon_A$ (the bias
$\mp\delta$ plus the common dimer-site shift $\Delta'$) and propagator
$G=(E-H_{QQ})^{-1}$, with
$\det=(E-\varepsilon_B)(E-\varepsilon_A)-\gamma_1^{2}$.  The two
$\gamma_0\otimes\gamma_4$ cross paths that connect the low-energy sites
through it ($A_1\!\to\!B_1\!\to\!B_2$ and $A_1\!\to\!A_2\!\to\!B_2$)
contribute
\begin{equation}
-\gamma_0\gamma_4 f^{2}\left[G_{BB}+G_{AA}\right]
=-\gamma_0\gamma_4 f^{2}\,\frac{2(E-\Delta')}{\det},
\end{equation}
because the opposite bias contributions to $\varepsilon_A$ and
$\varepsilon_B$ cancel.  The sum vanishes at $E=\Delta'$ for any bias, and
its $E$-odd remainder cancels between the valence and conduction evaluations,
which enter the interband dipole symmetrically.  What survives is second
order:\ $\gamma_0^{2}\to\gamma_0^{2}+\gamma_4^{2}$, a $(\Delta'/\gamma_1)^{2}$
dressing of the dimer propagator, and a mixed term
$\propto\gamma_4\Delta'/(\gamma_0\gamma_1)$.

Keeping the quadratic term of the
structure factor,
$f(K{+}\bm q)\simeq-\tfrac{\sqrt3 a}{2}\xi^{*}+\tfrac{a^{2}}{8}\xi^{2}$ (the monolayer element $-\gamma_0 f$ is the conjugate of the one that carries $\xi$, hence the $\xi^{*}$), the
gapped monolayer acquires the bilayer structure with reversed roles
(chirality $\xi^{*}$, warping $\xi^{2}$), and the balance that places the
bilayer vortex at $q_{\rm opt}$ [Eq.~(\ref{eq:qopt})] formally predicts a
$w=-1$ zero at $|q|=2\sqrt3/a\approx1.4$~\AA$^{-1}$.  Both terms, however,
carry the single energy scale $\gamma_0$, so this radius is fixed by the
lattice constant alone.  It lies outside the validity of the expansion, and
in the full lattice model the would-be zero is absorbed into the doubly dark
($w=0$) selection rule at $\Gamma$, leaving the valley strictly zero-free.

\section{The pinned charge at odd $N$}
\label{app:oddN}

This appendix derives the odd-$N$ pinned charge of Rule~4 and shows why
the bias cannot move it.

The monomials of $\bar h_N$ have degrees $N-s$, $s\le\lfloor N/2\rfloor$,
so the minimal degree is $(N+1)/2$ for odd $N$, attained only by
$\xi(\xi^{*})^{n}$, $n=(N-1)/2$; the bias enters the L\"owdin denominators
only and cannot generate lower-degree monomials.  Differentiating, with $\xi=q\,e^{i\theta}$ ($\theta$ the polar angle of $\bm q$),
\begin{multline}
\partial_{k_y}\!\left[\xi(\xi^{*})^{n}\right]
= i\!\left[(\xi^{*})^{n}-n\,\xi(\xi^{*})^{n-1}\right]\\
= i\,q^{n}e^{-i(n-1)\theta}\!\left[e^{-i\theta}-n\,e^{i\theta}\right],
\end{multline}
whose two coefficients descend from the same monomial: their ratio is
locked at $n$ by the differentiation itself.  For $n\ge2$ the second
component dominates everywhere on a small loop, giving winding
$-(n-1)+1=2-n=-(N-5)/2$; the same holds for the $x$ derivative
[$e^{-i\theta}+n\,e^{i\theta}$].  Only $n=1$ ($N=3$) is marginal.
Together with the even branch, $w_K=-(N-2)/2$, the two formulas form a
single arithmetic ladder, the odd values repeating the even ones three layers
later [$w_K(4)=w_K(7)=-1$, $w_K(6)=w_K(9)=-2$]; the sequence $-1,0,-2,-1$ for
$N=4$--$7$ is two interleaved ladders, not a parity alternation.

We have checked these windings numerically on small loops around $K$:\ they are $w_K=0,-1,-2$ for
$N=5,7,9$, independent of $U$, and the total near-$K$ charge recloses at
$N-1$ once the loop encloses the satellite ring.

\section{Zeros beyond the constellation}
\label{app:census}

This appendix supports Sec.~\ref{sec:counting}:\ it derives the
lattice-scale zeros, gives the exact locus on which they sit and their
census, and describes how the staircase and the direct winding $W$ are
computed.

Away from $K$ the two interlayer channels
carry different momentum factors.  Projected on the monolayer doublets, whose
eigenvectors carry the sublattice phase $f^{*}/|f|$, the dimer bond enters as
$\gamma_1 f$ and the skew bond, which collects its own momentum factor as
well, as $\gamma_3 f^{*2}$.  The effective interlayer coupling is their
difference,
\begin{equation}
t(\vk)=\frac{\gamma_1 f-\gamma_3 f^{*2}}{2|f|},
\label{eq:tk}
\end{equation}
and optical zeros appear where the two channels interfere
destructively, $t=0$.  Separating modulus and phase, this requires
\begin{equation}
|f|=\frac{\gamma_1}{\gamma_3}
\qquad\text{and}\qquad
3\arg f\equiv0 \ \ (\mathrm{mod}\ 2\pi),
\label{eq:seed}
\end{equation}
one radius and \emph{three} discrete angles per valley, which is why the
bilayer family appears as an interference triplet rather than in arbitrary
number.

With $\gamma_4=0$ every momentum-dependent bond
of the bilayer carries $f$ or $f^{*}$, so
$\partial_a H=(\partial_a f)V_1+(\partial_a f)^{*}V_1^{\dagger}$ with the
constant bond matrix
$V_1=-\gamma_0(|A_1\rangle\langle B_1|+|A_2\rangle\langle B_2|)
-\gamma_3|B_2\rangle\langle A_1|$.  An optical zero of any linear
polarization then requires $|M_1|=|M_3|$, with
$M_{1,3}=\langle v|V_1^{(\dagger)}|c\rangle$:\ one real condition that
defines a closed, trigonally warped ring around each valley, on which a
polarization-dependent phase condition selects isolated points.  The seeds of
Eq.~(\ref{eq:seed}) are the leading approximation to this ring, and every
lattice-scale zero lies on it.

For AB in $y$ polarization, the $K$ valley hosts,
besides the constellation, the interference triplet at
$|q|\simeq0.41$~\AA$^{-1}$ (angles $180^{\circ}$ and $\pm56^{\circ}$, each
$w=-1$) and a further pair at $\pm83^{\circ}$, which step $\chi$ from $-1$
to $-4$ and then to $-6$ [Fig.~\ref{fig:summary}(b)].  With $w_\Gamma=0$, the
balance of Sec.~\ref{sec:counting} closes.  In ABC a
six-zero interference shell and the $\pm81^{\circ}$ pair (eight zeros in all) on the same ring take $\chi$ from $+2$ to $+8$ and then to $+10$
[Fig.~\ref{fig:summary}(d)], and in the rhombohedral (ABCA) tetralayer the interference
shell and the $\pm171^{\circ}$ pair take it from $+3$ through $+6$ and $+12$ to
$+14$ (not shown).

These constellations are mirror symmetric but not $C_3$ symmetric, and the asymmetry is not an artifact of the zero finder.  The operations that map a valley onto itself are the rotations $C_3$ about $K$ and the antiunitary product of the vertical mirror with time reversal, which sends $q_y\to-q_y$ and leaves $|v^{y}_{vc}|$ and $|v^{x}_{vc}|$ invariant; this product is what produces the $\pm\theta$ pairs.  $C_3$, by contrast, rotates $v^{x}_{vc}$ and $v^{y}_{vc}$ into each other as the two components of a vector, so the zeros of a single Cartesian component need not form $C_3$ multiplets:\ a rotation by $120^{\circ}$ maps the $y$ constellation onto that of the linear polarization at $210^{\circ}$, not onto itself (we have checked this numerically).  This is why the interference triplet sits at $\pm56^{\circ}$ rather than $\pm60^{\circ}$, and why the angle of the off-seed pair is not fixed by symmetry:\ at fixed $\gamma_1/\gamma_3$ it drifts smoothly with the bias, from $\pm81.5^{\circ}$ at $\delta=0.02$~eV to $\pm87.5^{\circ}$ at $\delta=0.5$~eV in AB, and at fixed bias it moves with $\gamma_1/\gamma_3$, from $\pm77^{\circ}$ at $\gamma_1/\gamma_3=1.9$ to $\pm84^{\circ}$ at $0.8$.  The combinations that are $C_3$ eigenvectors are the circular components $v^{\pm}_{vc}=v^{x}_{vc}\pm iv^{y}_{vc}$, which acquire only a phase under $C_3$; their lattice-scale zeros form exact triplets at $60^{\circ}$, $180^{\circ}$, and $-60^{\circ}$, with radii $0.44$ and $0.37$~\AA$^{-1}$ for $\sigma^{+}$ and $\sigma^{-}$ in AB at $\delta=0.15$~eV.
included.

The per-zero charges are
gauge-invariant circulations on small loops.  For the disk character
$\chi=\nu-F$ [Eq.~(\ref{eq:character})], $\nu$ is the circulation of the
shift-vector field $(R^{x,y},R^{y,y})$ on circles of radius $\rho$, with dark
points excluded, and $F$ is the unwrapped parallel-transport phase of the two
bands.  The direct winding $W$ shown alongside uses instead a fixed reference
spinor.  It is well defined only while the overlaps with that reference stay away
from zero, which fails at lattice-scale radii, so the figures show $W$ only
where it is reliable, and only $\chi$ is used at the shells.

\begin{figure}
\includegraphics[width=\columnwidth]{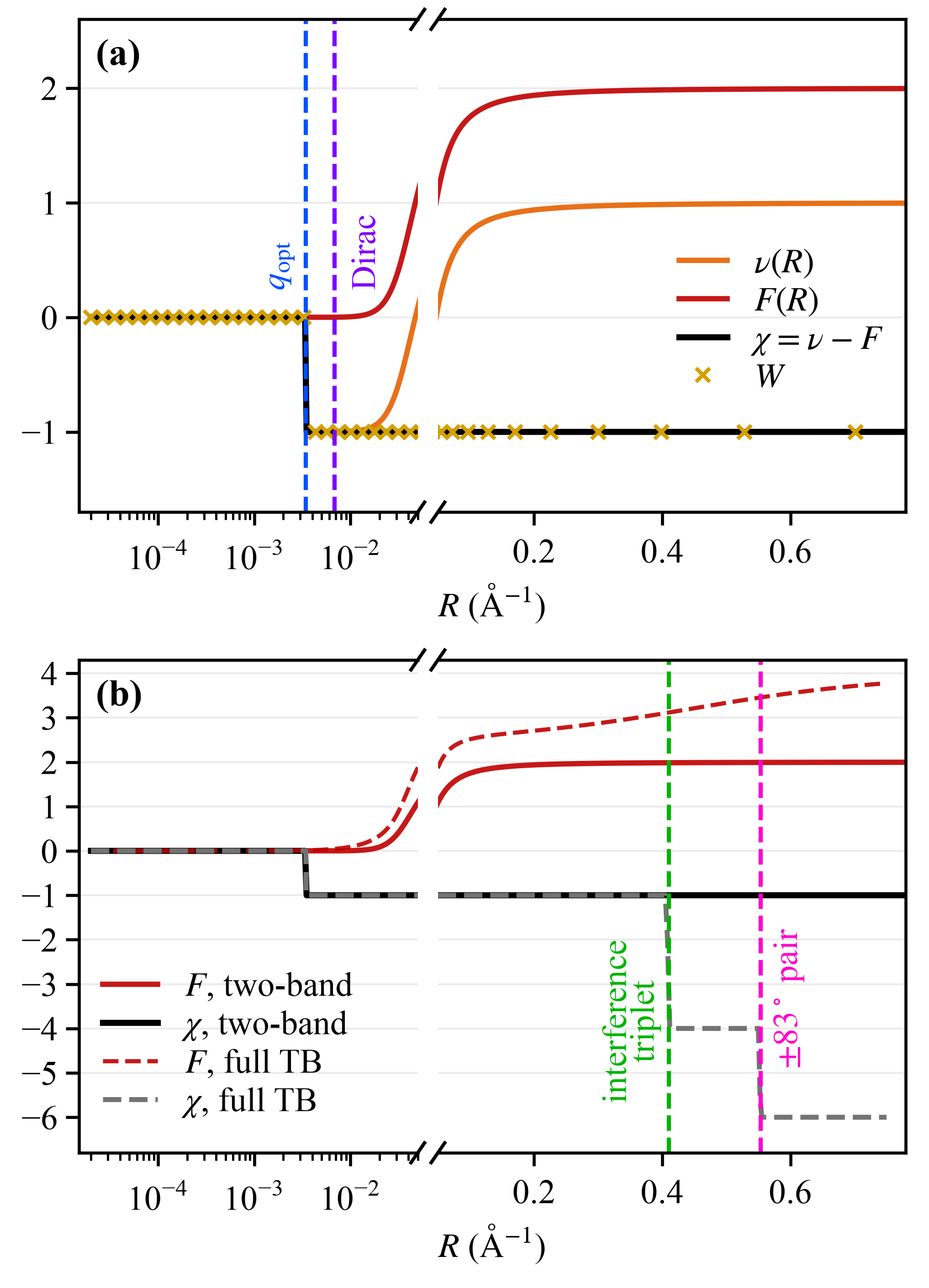}
\caption{Two-band effective model of the AB bilayer
($\gamma_3=0.38$~eV, $\delta=0.15$~eV; Appendix~\ref{app:census}), in the broken-axis format of
Fig.~\ref{fig:summary}.  (a)~$\nu$, $F$, $\chi$, and $W$ of the two-band
model; vertical lines mark $q_{\rm opt}$ and the satellite-Dirac radius $q_{\rm D}$ (labeled ``Dirac'').
(b)~$\chi$ and $F$ of the two-band model (solid) and of the full
$4\times4$ TB (dashed); vertical lines mark the interference
triplet and the $\pm83^{\circ}$ pair of the full model.  The disk radius $\rho$ is labeled $R$ in the figure axes.}
\label{fig:twoband}
\end{figure}

Dropping the dimer spectators reduces the path
expansion to the effective bilayer model
$h(\xi)=v_3\xi-(v_0^{2}/\gamma_1)\xi^{*2}$, written for the conjugate element $\bar h_2^{*}$ of Eq.~(\ref{eq:hN}), hence the chiral term $\xi^{*2}$.  Figure~\ref{fig:twoband}
compares its staircase with that of the full TB.  The two agree
near $K$, where the constellation lives, while the lattice-scale steps and the
nonquantized growth of $F$ belong to the full model alone.

\section{Derivation of the linear law $|\sigma|_{\rm pk}\propto N$}
\label{app:Nlaw}

This appendix derives statements (i)--(iii) of Sec.~\ref{sec:Nlaw}.

The chiral ($s{=}0$) term of Eq.~(\ref{eq:hN}) is
the monomial $\bar h_N^{(0)}=(v_0^{N}/\gamma_1^{N-1})\,\xi^{N}$, for
which homogeneity of degree $N$ gives the identity
\begin{equation}
\bigl|\partial_q\bar h_N^{(0)}\bigr|=\frac{N}{q}\,
\bigl|\bar h_N^{(0)}\bigr|.
\label{eq:euler}
\end{equation}
On the band-edge ring the off-diagonal
element balances the bias, $|\bar h_N(q_{\rm edge})|\simeq U$ (this is
what defines the band-edge ring), and the interband velocity is
dominated by $\partial_q\bar h_N$, so
\begin{equation}
\langle|v_{cv}|\rangle\simeq c_1\,\frac{N\,U}{q_{\rm edge}},
\label{eq:vring}
\end{equation}
with $c_1$ an $O(1)$ projection factor.  Equation~(\ref{eq:euler}) is
exact only for the chiral monomial, but $q_{\rm edge}$ lies above the
chiral--warping crossover ${\sim}v_3\gamma_1/v_0^{2}$ at every $N$, and $c_1$
drifts by only ${\sim}20\%$ while $N$ triples.

Along each direction
$\theta$ the transition energy is minimal on the ring,
$\omega_{cv}\simeq\omega_{\rm edge}(\theta)+\tfrac12\omega''(\theta)\,
[q-q_{\rm edge}(\theta)]^{2}$, where $q_{\rm edge}(\theta)$, of order $q_{\rm edge}$, is the ring
radius along $\theta$ and $\omega_{\rm edge}(\theta)$ the transition energy there.  When the broadened radial window
$\sqrt{2\eta/\omega''}$ is narrow compared with $q_{\rm edge}$, the measure may be
replaced by $q\,dq\to q_{\rm edge}\,dq$, and the broadened joint density of states
becomes
\begin{equation}
\mathcal{J}(\omega)=\int_0^{2\pi}\!\!d\theta\;
\frac{q_{\rm edge}(\theta)}{\sqrt{\eta\,\omega''(\theta)}}\;
\Phi\!\left(\frac{\omega-\omega_{\rm edge}(\theta)}{\eta}\right),
\label{eq:Wring}
\end{equation}
with the universal broadened van~Hove profile
$\Phi(z)=(2\pi)^{-1/2}\int dv\,e^{-(z-v^{2}/2)^{2}/2}$, of order one near
the edge.  The linear law uses a single property of this result, the
circumference factor $\mathcal{J}\propto q_{\rm edge}$, which cancels the
$1/q_{\rm edge}$ of Eq.~(\ref{eq:vring}) in the product $\langle I\rangle \mathcal{J}$;
the factor $1/\sqrt{\omega''\eta}$ belongs to the residual drift.

The velocity operator is layer-local
(intralayer $v_0$ and skew $v_3$ blocks only), and the band-edge states
reside on the outer sites $A_1/B_N$; $\partial H/\partial k$ therefore
couples them only to the outermost dimer states, at energies
${\pm}\gamma_1$, independently of the number of intervening dimers.  The
virtual sum thus contains $O(1)$ effective channels at every $N$,
\begin{equation}
\langle|T_{\rm virt}|\rangle\sim v_0^{2}/\gamma_1,
\label{eq:Tvirt}
\end{equation}
consistent with the computed ring average,
$\langle|T_{\rm virt}|\rangle\gamma_1/v_0^{2}\simeq1.1$--$1.4$ with no trend
for $N=2$--$6$.

Combining the three statements yields Eq.~(\ref{eq:NlawMain}).

\bibliography{refs}

\end{document}